# Light Trapping by Non-Hermitian Thin Films

Lina Grineviciute[1*], Ignas Lukosiunas[2], Julianija Nikitina[1], Algirdas Selskis[1], Indre Meskelaite[2], Darius Gailevicius[2], and Kestutis Staliunas[2,3,4*]

[1] Center for Physical Sciences and Technology, Savanoriu Ave. 231, LT-02300, Vilnius, Lithuania
[2] Vilnius University, Faculty of Physics, Laser Research Center, Sauletekio Ave. 10, Vilnius, Lithuania
[3] ICREA, Passeig Lluís Companys 23, 08010, Barcelona, Spain
[4] UPC, Dep. de Fisica, Rambla Sant Nebridi 22, 08222, Terrassa (Barcelona), Spain

E-mails: lina.grineviciute@ftmc.lt, kestutis.staliunas@icrea.cat

## Abstract

One of the exceptional features of non-Hermitian systems is the unidirectional wave interactions. Simultaneous modulation of the real and the imaginary part of the interaction potentials (of the refractive index and the gain/loss in the case of optical systems) can result in unequal coupling coefficients between the fields of different parts of the system. The unidirectional coupling can also be arranged not only between the internal fields of the system but also between internal fields and external radiation. At a particular (exceptional) point, the situation can be achieved, that the external radiation is efficiently coupled into the system, but the internal radiation cannot escape backwards. In this way, the incident radiation can be trapped inside the non-Hermitian system and, eventually, can be efficiently absorbed there.
We realize this idea in non-Hermitically modulated thin films. The modulation consists of a Hermitian part – the periodic corrugation of the surfaces of a thin film, and a non-Hermitian part – the modulation of losses along the film. We prove numerically and demonstrate experimentally that the incident radiation, coupled with such a non-Hermitian thin film, is unidirectionally trapped into a planar mode of the film, does not escape from the film (or escape weakly due to experimental imperfections), and is efficiently absorbed there.

Non-Hermitian potentials (the Parity-Time (PT) -symmetric potentials[1] being a class of it) offer unidirectional wave propagation-absorption effects, such as unidirectional invisibility[2–4] and other unidirectional effects[5–12]. The unidirectionality effects often can be interpreted by asymmetric coupling between the radiation modes of the system, or in simplest case between two plane waves of the same frequency but with different propagation wavevectors $\vec{k_1}$ and $\vec{k_2}$ ($\vec{k_1} - \vec{k_2} = \vec{q}$). If the electric susceptibility modulation of the coupler contains the Hermitian component $\cos(\vec{q}\vec{r}) = (e^{i\vec{q}\vec{r}} + e^{-i\vec{q}\vec{r}})/2$, then this component equally couples the wave $\vec{k_1}$ with the wave $\vec{k_2}$, and vice versa. Both exponents $e^{i\vec{q}\vec{r}}$ and $e^{-i\vec{q}\vec{r}}$ are equally participating in the Hermitian coupling potential, ensuring symmetric coupling between the two waves. This result rigorously follows, for instance, from the first Born approximation[13,14], where the spatial spectrum of a scattered field is proportional to the convolution of the spectrum of the incident wave and of the material susceptibility. However, if the susceptibility modulation contains only one exponential part, let us say, the $e^{i\vec{q}\vec{r}}$, i.e., only just a "half" of a cosine, then this would result in the wave $\vec{k_2}$ efficiently coupled to the wave $\vec{k_1}$, but not vice versa. The missing modulation component $e^{-i\vec{q}\vec{r}}$ in the latter case, breaks the symmetry of interaction. The unidirectional coupling potential $e^{i\vec{q}\vec{r}} = \cos(\vec{q}\vec{r}) + i\sin(\vec{q}\vec{r})$, however, consists of the real and imaginary parts, which means that not only the refractive index must be modulated as $\cos(\vec{q}\vec{r})$, but also the gain and/or loss must be modulated as well as $\sin(\vec{q}\vec{r})$.

The unidirectional coupling can result in the trapping of the radiation incident into a non-Hermitian system. In the ideal case of perfectly matched refractive index and gain/loss grating profiles $\varepsilon(\vec{r}) \sim \cos(\vec{q}\vec{r}) + i\sin(\vec{q}\vec{r})$, (equal amplitudes of modulation quadratures, and exactly the $\pi/2$ phase shift), the incident radiation efficiently enters into the system but cannot escape. This is an analog of the "black hole" for light. This can have profound practical consequences, since the photons, trapped into a system with losses without excite, should experience full absorption.

The physical situation is depicted in Fig.1.a. The incident radiation, at a near-to-normal incidence angle, couples with the planar-guided modes of the thin film due to the periodic modulation of one (or both) of its surfaces. A modulation of a surface of dielectric material provides the Hermitian part of the coupling. The incident radiation couples equally into the right- and left-propagating planar modes. The situation is different if the periodic absorption function is added to the Hermitian part of the potential. This can be considered as the "imaginary" part of the potential. If both, the Hermitian and non-Hermitian quadratures of modulations, are equal in amplitude and $\pi/2$ phase shifted one with respect to another, the incident radiation couples to the right-propagating mode only, but not to the left-propagating mode. Moreover, the right propagating mode does not couple back to the incident (and transmitted or reflected) radiation.

The analytical and numerical study of this effect, as well as the experimental proof, is the main message of the present letter. Experimental proof of the asymmetric absorption was performed by two methods. An indirect proof was provided by measuring the reflection, $R$, and transmission, $T$, of the incident radiation, and estimating the absorption as $A = 1 - R - T$. A direct experimental proof was obtained by photothermal measurements: absorption results in an increase of the temperature of the sample, thus the temperature measurements allow to estimate the absorption.

In theoretical study we consider the main ingredients of the system phenomenologically. We assume the net dielectric constant varies periodically along the film:

$$\Delta\epsilon(x) = \frac{\Delta\epsilon_r}{2}\cos(qx) + \frac{i\Delta\epsilon_i}{2}(\sin(qx) + 1) =$$
$$\frac{\Delta\epsilon_r}{4}(e^{iqx} + e^{-iqx}) + \frac{\Delta\epsilon_i}{4}(e^{iqx} - e^{-iqx}) + \frac{i\Delta\epsilon_i}{2}. \quad (1)$$

This expression implies refractive index modulated as $\Delta\epsilon_r \cos(qx)/2$, and absorption consisting of a varying part $\Delta\epsilon_i \sin(qx)/2$, and a constant background part $\Delta\epsilon_i/2$. The background part is inevitable, since no gain but only losses is present in our system. For the reference structure we consider only the background part of the losses $\Delta\epsilon_i/2$. In general, the dielectric constant can be written as:

$$\Delta\epsilon(x) = \Delta\epsilon_+ e^{iqx} + \Delta\epsilon_- e^{-iqx} + i\Delta\epsilon_0, \quad (2)$$

with arbitrary real-valued $\Delta\epsilon_+$, $\Delta\epsilon_-$, $\Delta\epsilon_0$, which stand for right/left coupling and background losses, respectively. For the materials with losses only: $\Delta\epsilon_0 \geq \Delta\epsilon_+ + \Delta\epsilon_-$. For the materials with gain and loss: $\Delta\epsilon_0 \leq \Delta\epsilon_+ + \Delta\epsilon_-$, and $\Delta\epsilon_0$ can obtain small values $\Delta\epsilon_0 \to 0$. In the latter case asymmetries of the absorption can be very large. In our experiments, however, only the losses were possible, no gain, which imposes the restriction on the coefficients $\Delta\epsilon_0 \approx \Delta\epsilon_+ + \Delta\epsilon_-$.

In our analytical treatment (see Supplementary material), we split the fields into two parts: the Fabry-Perot (FP) part of the radiation, which is the radiation nearly normal to the film, $B$, and to the part corresponding to the planar waveguiding modes, nearly parallel to the film, $A$:

$$\frac{\partial A}{\partial t} = i\Delta\omega A + i\beta_1 B - \alpha A, \quad (3.a)$$

$$\frac{\partial B}{\partial t} = \frac{B_0\sqrt{f} - B}{\tau} - \alpha B + i\beta_2 A. \quad (3.b)$$

The Fabry-Perot field $B$ feeds from the incident wave $B_0$, with $\tau$ being the photon lifetime of the FP mode, and $f$ is the FP resonator finesse. The resonance for FP field is considered, as the general case, does not bring qualitative differences. The planar waveguiding component $A$, in general, is detuned from the resonance by $\Delta\omega = \omega - \omega_0 = (k - k_0)/c$, and experiences absorption $\alpha$. The absorption is restricted to $\Delta\epsilon_i \omega_0/4$ in the presence of the losses only, no gain, which is the case in our experiments. In the presence of gain the $\alpha$ can obtain arbitrary values, including negative ones for the gain dominating over losses. The complex-valued coupling coefficients $\beta_1$ and $\beta_2$ are no more complex conjugated, as in the Hermitian case, but can be arbitrary. In our case $\beta_{1,2} = (\Delta\epsilon_r + i\Delta\epsilon_i e^{\pm i\varphi})\omega_0/2$, where $\varphi$ is the phase shift between the real and imaginary modulation components.

The stationary solution of (3) gives the expression for the amplitude of the planar mode:

$$A = \frac{i\beta_1\sqrt{f}}{(\alpha - i\Delta\omega_A)(1 + \alpha\tau) + \beta_1\beta_2\tau}B_0. \quad (4)$$

The maximum amplitude of the planar wave (at resonance $\Delta\omega_A = 0$, and for weak losses $\alpha \ll \beta_{1,2}$) gives a simple expression: $A = i\sqrt{f}B_0/(\beta_2\tau)$, which predicts an unbounded increase in the intensity of the trapped field for vanishing back-coupling $\beta_2 \to 0$. In reality, the fields will be restricted by the losses $\alpha > 0$, or by the saturation of the gain $\alpha < 0$.

Next, we compare the absorption in the non-Hermitian sample with the absorption in the reference case, i.e., for the same amount of absorber distributed uniformly over the sample. In the ideal case (for ideal unidirectional coupling $\Delta\epsilon_i = \Delta\epsilon_r$, which means that there is no back coupling), $\beta_2 = 0$, $\beta_1 = \alpha$, the peak intensity absorption is given by: $\gamma(0) \approx 4\alpha f$ (peak absorption occurs at the resonance). In the reference case with the uniform absorption: $\Delta\epsilon_i = 0$, $\beta_1 = i\beta_2 = \alpha/2$, and the peak intensity absorption is: $\gamma(0) \approx 5\alpha f/2$, This means that the unidirectional coupling can exceed the absorption in the symmetric coupling case by a factor of $8/5$, i.e., almost twice. Note that we considered the condition $\alpha\tau \ll 1$.

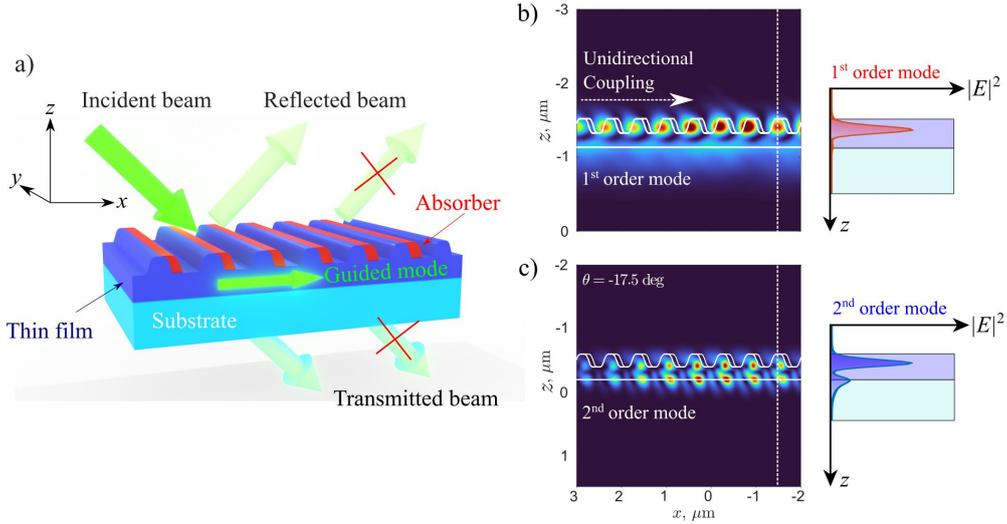

**Fig.1. a)** The scheme, where the incident wave is resonantly coupled with the guided mode of the film via periodic modulation of the surface of the film (modulated in the direction along the film $x$), with a period equal to the wavelength of the guided mode; The back coupling to the incident radiation modes is absent at a critical point, or weak in general. **b, c)** the field intensity profiles averaged along the transverse coordinate $x$ with calculated planar mode profiles, corresponding to the lowest and next higher waveguiding modes

In the presence of gain the field enhancement factor is larger than the dissipative limit $8/5$, as (4) indicates, and increases to infinity approaching the gain/loss balance point.

These analytical predictions were verified by a numerical simulation of the fields shown in the scheme of Fig.1.a. Numerical study was performed by the Rigorous Coupled Wave Analysis (RCWA) method[15,16], also checked by a finite difference (FDTD) method[17]. The parameters used in numerical 2D RCWA simulation were chosen to correspond to the experiment: the period of modulation $\Lambda_x = 0.625$ µm, modulation depth $h = 0.2$ µm, the film thickness $L = 0.5$ µm, refractive index of the substrate: $n_{sub} = 1.47$ (fused silica), the refractive index of the thin film material $n_{film} = 2.35$ ($TiO_2$), and the complex index of refraction of an absorber, $n_A = 2.4 - 2.3i$ (corresponds to nickel at 970 nm wavelength as used in experiments).

The numerical results are summarized in Fig.2, where the transmission, reflection and absorption maps in the plane of incidence angle and wavelength are shown for $s$ polarization. The reflection/transmission maps have a structure of tilted and crossing resonance lines in the (angle, wavelength) space. The right/left tilted resonance lines correspond to the excitation of left/right propagating planar modes.

Fig.2.b. shows the symmetrical coupling case for the reference sample, where the absorber layer is uniform. Evidently, the reflection/transmission of the incident radiation through the reference sample is symmetric with respect to the incidence angle.

An interesting observation is that for the non-Hermitian case, the reflection map remains symmetric. The intuitive interpretation is that at the resonances with left- and right-propagating waveguide modes the reflections are equal to $\beta_1 \alpha \beta_2$ and $\beta_2 \alpha \beta_1$ respectively (in-coupling->absorption->out-coupling), i.e., a commutative form with respect to in- and out-coupling coefficients $\beta_1$ and $\beta_2$. The transmission map is, however, asymmetric, indicating asymmetric losses. Asymmetric losses are evident from the observation that the left-right propagating modes are of different intensity, according to (3) in the analytical model, and not commutative with respect to in- and out-coupling coefficients $\beta_1$ and $\beta_2$.

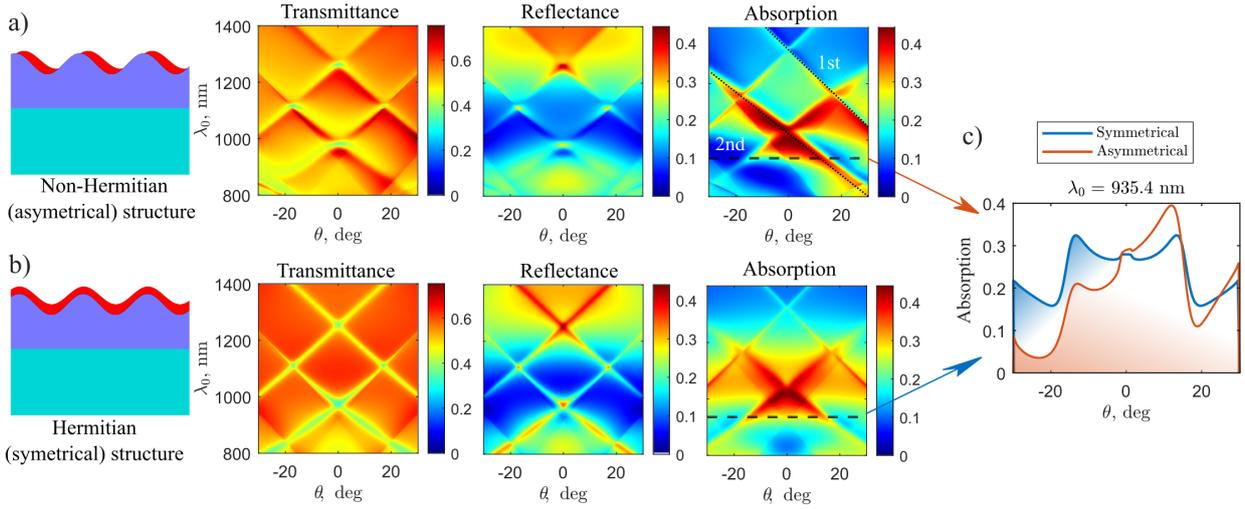

**Fig.2.** a) Non-Hermitian planar waveguide configuration, with periodically modulated surface and the absorption layer (red color) followed by the transmittance, reflectance and indirectly calculated absorption maps, as $A = 1 - R - T$, b) the reference configuration with uniformly distributed absorber along the modulated planar structure with transmission, reflection and absorption maps; c) The calculated angular absorption profiles for the scheme with asymmetrically (red) and symmetrically (blue) distributed losses (cross-sections of absorption maps for the fixed wavelength of 935.4 nm).

**Fabrication.** We fabricated the structure on a commercially available fused silica (FS) substrate, which was initially modulated by grooves of a rectangular shape. Attempts to deposit a uniform film on the rectangularly shaped substrate failed, therefore, we preprocessed the substrate (by ion-etching) to round the sharp edges of the grooves. Preprocessing is a purely technological issue, as the subsequent thin layer deposition is more efficient on the rounded substrate[18]. Next, layer deposition was performed via the ion beam sputtering method, where the titanium dioxide ($TiO_2$) layer was deposited. The surface of the film approximately reproduces the profile of the substrate, i.e., the film remains of a fixed thickness. Lastly, magnetron sputtering with the glancing angle deposition method was applied for asymmetric nickel (Ni) deposition. Technological description of such method can be found here[19,20]. The geometrical parameters and materials are as provided for numerical simulations, the only difference is that the substrate analyzed here does not have a flat interface. This, however, brings no qualitative differences with a single interface modulation considered in Fig.1,2. The fabricated structure as well as a Scanning Electron Microscope (SEM) image of its cross-section are presented in Fig.3.

In order to obtain non-Hermitian effects and induce asymmetrical absorption, a layer of nickel is deposited on the right-hand side of the guided mode resonance waveguide material. The distribution of chemical elements on the sample surface was studied by Energy Dispersive X-ray Spectroscopy and is presented by its X-ray lines intensities in Fig.3.c. The measurements show that indeed the absorber (nickel) is positioned approximately $\pi/2$ shifted from the background modulation maxima. The thickness of the titanium thin film is relatively uniform.

The absorption profiles of experimental part were first measured ***indirectly*** by measuring total reflection, $R$, and transmission, $T$, and calculating $A = 1 - R - T$. The total reflection and transmission are the summary of all reflection and transmission orders (not only the zero-reflection transmission order). Transmission and reflection maps for the fabricated sample were recorded by a spectrophotometer (provided in Supplementary material Fig.S2). Here, linearly polarized light was used for two perpendicular polarizations: $s$ and $p$, where $s$ polarization is parallel to the grating lines on the sample. The angle between the main plane of the grating and the detector was varied from -30° to 30° by steps of 1°.

Fig.4 shows the absorption spectra of the fabricated structure. An evident asymmetry of (angle, wavelength) map is visible for both $s$- and $p$- polarized inputs. The absorption associated with the right propagating modes (the left tilted resonance lines) is stronger than the absorption associated with the left propagating modes. Furthermore, comparison with the numerically calculated absorption maps are presented, showing a good correspondence.

In order to measure the asymmetric absorption ***directly***, we performed measurements via the photothermal spectroscopic approach[21]. If there are no other energy dissipation mechanisms, absorption results in heat release in the sample, resulting in increase of its temperature. This increase can be observed with a calorimeter [22] or by means of thermal imaging [23,24]. We measured the sample temperature dependence on the beam incidence angle for a fixed illumination time interval. The details of the experimental set-up are provided in Supplementary material (Fig. S3).

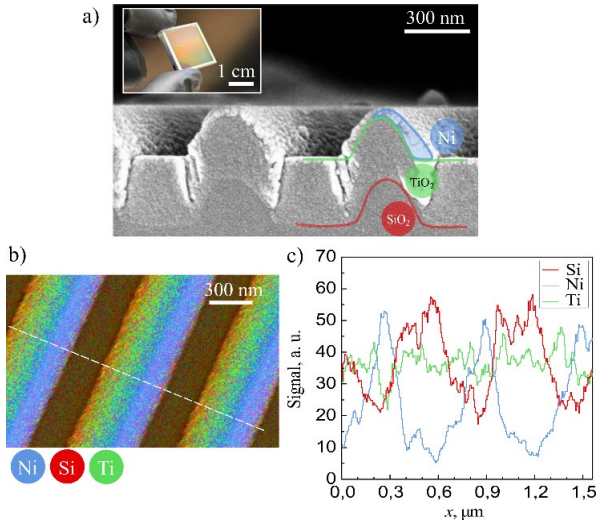

**Fig.3. a) Real image (inset) and cross-sectional SEM image; b) Energy Dispersive X-Ray Spectroscopy Analysis (sample top view) showing the distribution of different chemical elements within the surface layer, and c) chemical elements distribution on the sample surface on a line perpendicular to the grooves.**

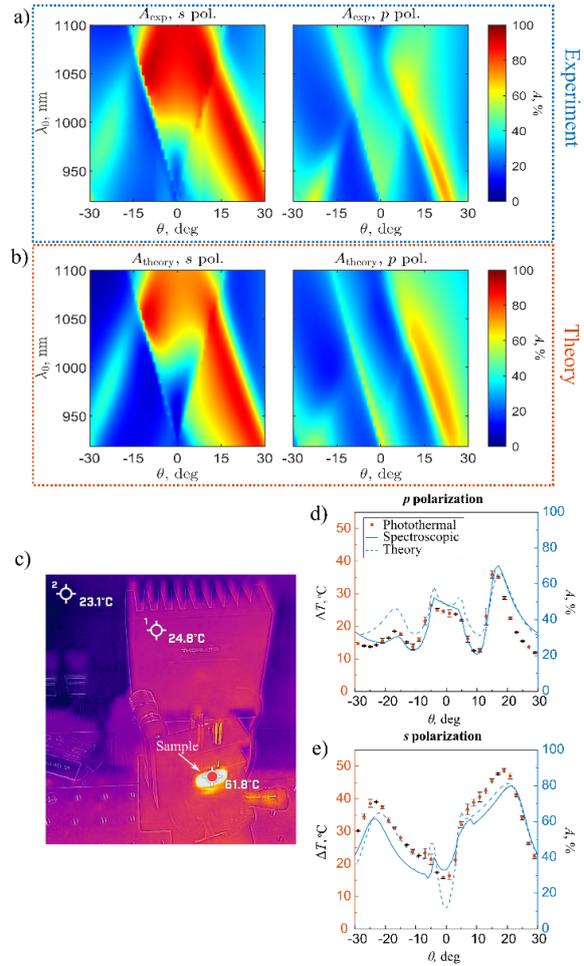

**Figure 4. a) absorption maps of the fabricated structure for *s* and *p* polarizations obtained by indirect measurements ($A = 1 - R - T$), together with b) numerically calculated absorption; c) a thermal image of the photothermal measurements set-up; d), e) photothermal measurements of the absorption of *p*- and *s*-polarized light ($\lambda$ = 970 nm), respectively, compared with the indirect spectroscopic measurements and numerical estimations.**

To irradiate the sample, we have used a collimated continuous wave laser diode beam of a central wavelength $\lambda$ = 970 nm with a full-width half maximum line thickness of 2.5 nm. At every angle, the sample was irradiated for 30 seconds with an average power of 2W. The increase of the temperature from the room temperature $T_0$ to $T_{max}(\theta)$ was observed. The change in temperature $\Delta T = T_{max}(\theta) - T_0$ was recorded using a thermal imager, which was arranged above the sample along the optical axis of the incident probe beam. Between each measurement the sample was allowed to cool down to room temperature. The obtained results are shown in Fig.4.d and Fig.4.e for *p* and *s* polarizations, respectively, along with the results of the indirect spectroscopic measurements and numerical estimations for comparison.

We could not calibrate the measured temperature increase with the absorption coefficient: the proportionality coefficient between these two quantities remains undefined. Therefore, the scales of the sample temperature change and absorption were fitted to maximum correspondence with the other (experimental indirect, and numerical) curves. We note that the shapes of the curves of all three cases are very similar, thus proving the idea of unidirectional trapping.

In **conclusion,** we have predicted and demonstrated a new effect, light trapping in a system, due to unidirectional coupling with the internal radiation modes of the system. By analytic/numerical/experimental study we show that this leads to enhanced absorption of the trapped field, compared to the case with symmetric coupling. In perspective, such unconventional couplers could be used as the first truly unidirectional devices for advanced light trapping, for static solar panels, for photodetectors, and nonreflecting optics. With the simplicity of the current approach requiring just a grating and an asymmetrical lossy coating, we expect numerous technological applications in the future.

# References

1. El-Ganainy, R., Khajavikhan, M., Christodoulides, D. N. & Ozdemir, S. K. The dawn of non-Hermitian optics. Commun Phys 2, 37 (2019).
2. Longhi, S. Parity-time symmetry meets photonics: A new twist in non-Hermitian optics. EPL (Europhysics Letters) 120, 64001 (2017).
3. Horsley, S. A. R., Artoni, M. & La Rocca, G. C. Spatial Kramers–Kronig relations and the reflection of waves. Nat Photonics 9, 436–439 (2015).
4. Huang, Y., Shen, Y., Min, C., Fan, S. & Veronis, G. Unidirectional reflectionless light propagation at exceptional points. Nanophotonics 6, 977–996 (2017).
5. Coppolaro, M. et al. Extreme-Parameter Non-Hermitian Dielectric Metamaterials. ACS Photonics 7, 2578–2588 (2020).
6. Ahmed, W. W., Herrero, R., Botey, M., Wu, Y. & Staliunas, K. Restricted Hilbert Transform for Non-Hermitian Management of Fields. Phys Rev Appl 14, 044010 (2020).
7. Foland, S. J. & Lee, J.-B. A highly-compliant asymmetric 2D guided-mode resonance sensor for simultaneous measurement of dual-axis strain. in 2013 IEEE 26th International Conference on Micro Electro Mechanical Systems (MEMS) 665–668 (IEEE, 2013). doi:10.1109/MEMSYS.2013.6474329.
8. Feng, L., El-Ganainy, R. & Ge, L. Non-Hermitian photonics based on parity–time symmetry. Nat Photonics 11, 752–762 (2017).
9. Grineviciute, L., Nikitina, J., Babayigit, C. & Staliunas, K. Fano-like resonances in nanostructured thin films for spatial filtering. Appl Phys Lett 118, 131114 (2021).
10. Liu, T. et al. Dislocated Double-Layer Metal Gratings: An Efficient Unidirectional Coupler. Nano Lett 14, 3848–3854 (2014).
11. Jia, Y., Yan, Y., Kesava, S. V., Gomez, E. D. & Giebink, N. C. Passive Parity-Time Symmetry in Organic Thin Film Waveguides. ACS Photonics 2, 319–325 (2015).
12. Aguilera-Rojas, P.J., Alfaro-Bittner, K., Clerc, M.G. et al. Nonlinear wave propagation in a bistable optical chain with nonreciprocal coupling. Commun Phys 7, 195 (2024)
13. Born, M. Quantenmechanik der Stoßvorgänge. Zeitschrift für Physik 38, 803–827 (1926).
14. Landau, L. D. & Lifshitz, E. M. Quantum Mechanics: Non-Relativistic Theory. vol. 3 (Elsevier, 2013).
15. Moharam, M. G., Gaylord, T. K., Grann, E. B. & Pommet, D. A. Formulation for stable and efficient implementation of the rigorous coupled-wave analysis of binary gratings. Journal of the Optical Society of America A 12, 1068 (1995).
16. Moharam, M. G., Gaylord, T. K., Pommet, D. A. & Grann, E. B. Stable implementation of the rigorous coupled-wave analysis for surface-relief gratings: enhanced transmittance matrix approach. Journal of the Optical Society of America A 12, 1077 (1995).
17. Rumpf, R. C., Garcia, C. R., Berry, E. A. & Barton, J. H. Finite-Difference Frequency-Domain Algorithm for Modeling Electromagnetic Scattering from General Anisotropic Objects. Progress In Electromagnetics Research B 61, 55–67 (2014).
18. Grineviciute, L. et al. Nanostructured Multilayer Coatings for Spatial Filtering. Adv Opt Mater 9, 2001730 (2021).
19. Hawkeye, M. M., Taschuk, M. T. & Brett, M. J. Glancing Angle Deposition of Thin Films. (Wiley, 2014). doi:10.1002/9781118847510.
20. Grineviciute, L. et al. Optical anisotropy of glancing angle deposited thin films on nano-patterned substrates. Opt Mater Express 12, 1281 (2022).
21. Tam, A. C. Overview Of Photothermal Spectroscopy. in LEOS '90. Conference Proceedings IEEE Lasers and Electro-Optics Society 1990 Annual Meeting vols 1990-Novem 154–157 (IEEE, 1990).
22. Decker, D. L. & Temple, P. A. The design and operation of a precise, high sensitivity adiabatic laser calorimeter for window and mirror material evaluation. Glass, AJ; Guenther, AH, ed. in Proceedings of the 9th annual symposium on optical materials for high power lasers 4–6 (1977).
23. Chen, P. et al. Facile syntheses of conjugated polymers for photothermal tumour therapy. Nat Commun 10, 1192 (2019).
24. Jin, Z. et al. Coordination-induced exfoliation to monolayer Bi-anchored MnB2 nanosheets for multimodal imaging-guided photothermal therapy of cancer. Theranostics 10, 1861–1872 (2020).

## Acknowledgements

**L.G.** and **J.N.** were supported by the PerFIN project from the Research Council of Lithuania (LMTLT), agreement No. S-MIP-22-80. **K.S.** was supported by Spanish Ministry of Science, Innovation and Universities (MICINN) under the project PID2022-138321NB-C21. **D.G.** and **I.M.** acknowledge support by the framework of the "Universities` Excellence Initiative" program by the Ministry of Education, Science and Sports of the Republic of Lithuania under the agreement with the Research Council of Lithuania (project No. S-A-UEI-23-6).

## Authors contribution

K.S. proposed the idea, supervised the research work, and performed analytical study. I.L. performed numerical simulations. L.G. and J.N. carried out experiments, sample fabrication, coating depositions, and spectroscopic measurements. A.S. performed SEM and EDX measurements. D.G. and I.M. carried out photothermal measurements. All authors contributed to the discussion of the results and to the edits of the manuscript.

## Competing interests

The authors declare no competing interests.